\documentclass{iau}
\usepackage{graphicx}
\usepackage{natbib}

\def\degree{\hbox{$^\circ$}}
\def\kms{\hbox{km$\;$s$^{-1}$}}

\title[Rayleigh–Taylor instability prominences]
{Rayleigh–Taylor instability in partially ionized prominence plasma}

\author[E. Khomenko, A. D\'{\i}az, A. de Vicente, M. Collados \& M. Luna]   %% give here short author list %%
{E. Khomenko$^{1,2}$, A. D\'{\i}az$^{1,2}$, A. de Vicente$^{1,2}$, M. Collados$^{1,2}$ \and M. Luna$^{1,2}$}

\affiliation{$^1$Instituto de Astrof\'{\i}sica de Canarias, 38205 La Laguna,
Tenerife, Spain \\ email: {\tt khomenko@iac.es} \\[\affilskip]
$^2$Departamento de Astrof\'{\i}sica, Universidad de La
Laguna, 38205, La Laguna, Tenerife, Spain}

\pubyear{2013}
\volume{xxx}  %% insert here IAU Symposium No.
\pagerange{1---4}
\jname{Title of your IAU Symposium}
\editors{A.C. Editor, B.D. Editor \& C.E. Editor, eds.}
\begin{document}

\maketitle

\begin{abstract}
We study Rayleigh--Taylor instability (RTI) at the coronal--prominence
boundary by means of 2.5D numerical simulations in a single-fluid MHD
approach including a generalized Ohm's law. The initial configuration
includes a homogeneous magnetic field forming an angle with the direction in
which the plasma is perturbed.  For each field inclination we compare two
simulations, one for the pure MHD case, and one including the ambipolar
diffusion in the Ohm's law, otherwise identical. We find that the
configuration containing neutral atoms is always unstable. The growth rate of
the small-scale modes in the non-linear regime is larger than in the purely
MHD case.
\end{abstract}

Solar prominences are blocks of cool and dense chromospheric plasma remaining
stable for days, or even weeks, in the solar corona. Despite the global
stability, prominences are extremely dynamic at small scales, showing a
variety of shapes, moving with vertical and horizontal threads.
\citet{Berger2010} find large-scale 20--50 Mm arches, expanding from
underlying corona into the prominences. At the top of these arches, at the
border between the corona and the prominence, there are observed dark
turbulent upflowing channels of 4-6 Mm maximum width with a profile typical
for the Rayleigh--Taylor (RT) and Kelvin--Helmholtz (KH) instabilities
\citep{Berger2010, Ryutova+etal2010}. The upflows rise up to 50--15 Mm, with
an average speed of 13--17 \kms, decreasing at the end. Lifetimes of the
plumes are about 300--1000 sec. From the theoretical point of view, the
existence of the instabilities at the interface between the prominence and
the corona is easily explained since the two media have clearly different
densities, temperatures and relative velocities. Recent numerical simulations
of the RTI by \citet{Hillier+etal2012a}, including a rising buoyant tube in a
Kippenhahn--Schl\"{u}ter prominence model show a good agreement with
observations. Nevertheless, prominence plasma is expected to be only
partially ionized. The presence of a large number of neutrals must affect the
overall dynamics of the plasma. Linear theory of the Rayleigh--Taylor and
Kelvin--Helmholtz instabilities in the partially ionized plasma has been
recently developed by \citet{Soler+etal2012, Diaz+etal2013} showing that
there is no critical wavelength as in the purely mhd case
($\lambda_c=B_0^2\cos^2\theta/(\rho_2-\rho_1)/g$), and perturbations in all
the wavelength range are always unstable. The aim of our work is to model the
dynamics of the Rayleigh--Taylor instability in the partially ionized
prominence plasma in the non-linear regime.

%%%%%%%%%%%%%%%%%%%%%%%%%%%%%%%%%%%%%%%%%%%%%%%%%%%%%%%%%%%%%%%%
\begin{figure*}
\center
\includegraphics[width=16cm]{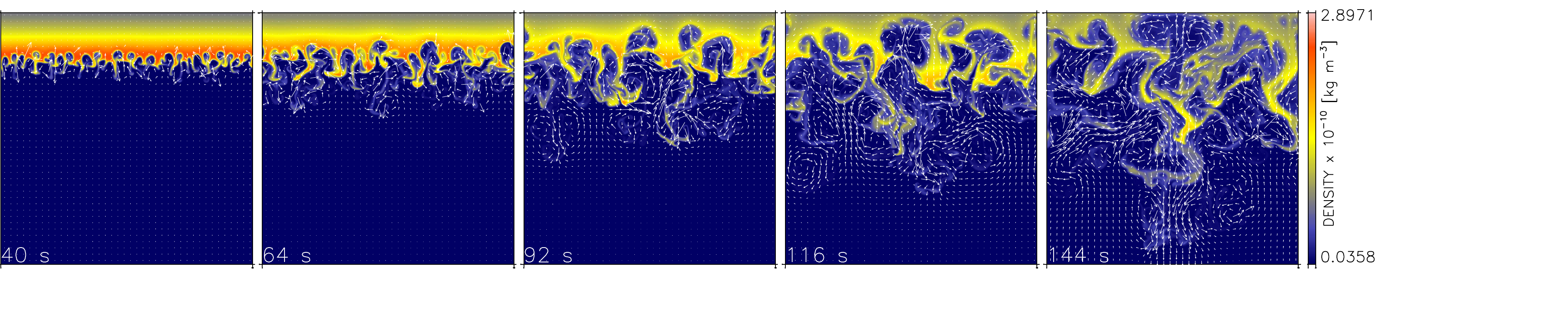}
\includegraphics[width=16cm]{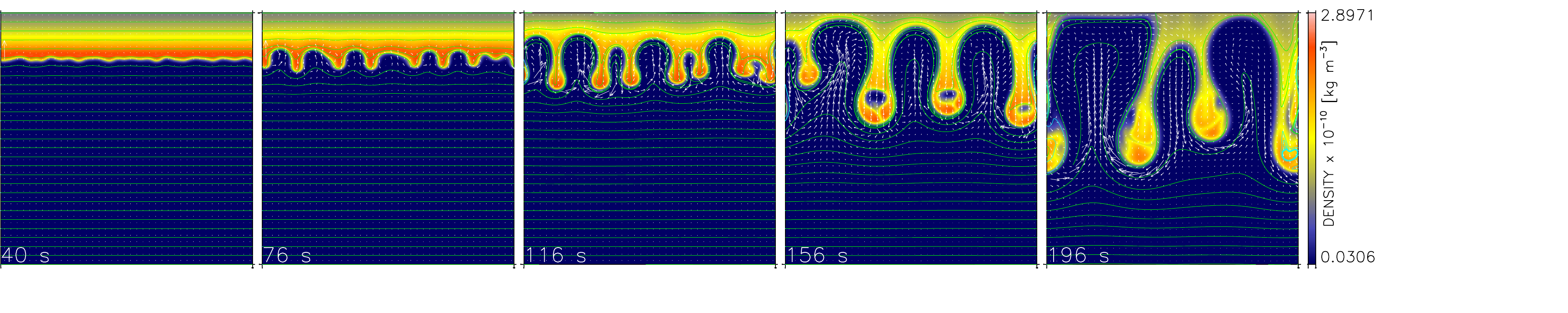}
\includegraphics[width=16cm]{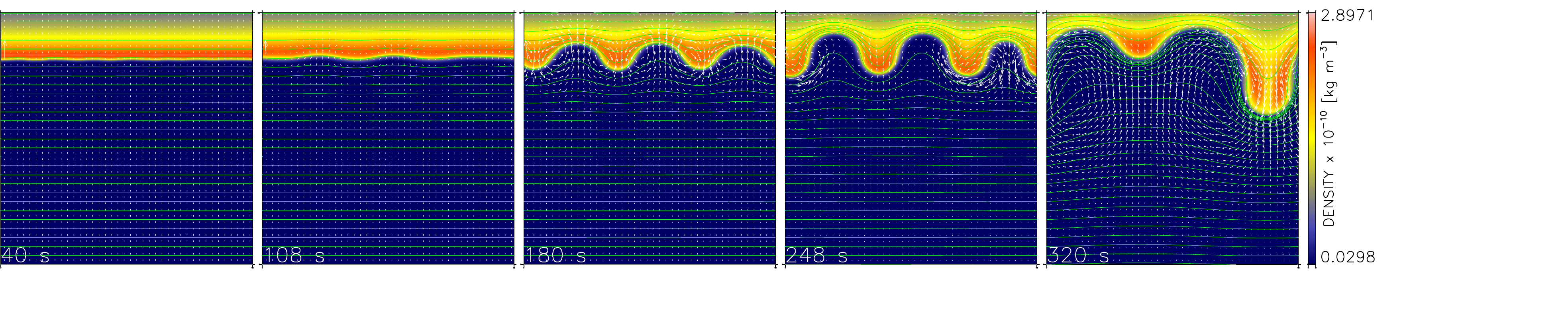}
\caption{Top: Time evolution of density in the ``ambipolar'' simulation with
$\theta=90$\degree. Middle: same for the simulation with $\theta=89$\degree.
Green lines are projections of magnetic field lines into $XZ$ perturbation
plane. Bottom: same for the simulation with
$\theta=88$\degree.}\label{fig:tevol}
\end{figure*}
%%%%%%%%%%%%%%%%%%%%%%%%%%%%%%%%%%%%%%%%%%%%%%%%%%%%%%%%%%%%%%%%

We simulate a small portion of the interface between prominence and corona,
of the size of 1$\times$1 Mm. The initial stratification of the pressure and
density is in hydrostatic equilibrium for a given temperature. The
equilibrium magnetic field, $\vec{B}_0$, is homogeneous and does not
influence the force balance. The plasma is perturbed in the $XZ$ plane, the
magnetic field vector is initially in $XY$ plane, making an angle with $X$
axis. We consider 3 simulation runs with $\vec{B}_0$ at $\theta=90$\degree\
to $X$ axis (i.e. normal to the $XZ$ plane), $\theta=89$\degree\ and
$88$\degree. The following parameters of the equilibrium configuration were
taken: $T_{\rm cor}=400$ kK, $T_{\rm chrom}=5$ kK; $\rho_{\rm
cor}=3.7\times10^{-12}$ kg m$^{-3}$, $\rho_{\rm chrom}=2.9\times10^{-10}$ kg
m$^{-3}$, and $B_0=10$ G. This provides chromospheric neutral fraction of
$\rho_n/\rho=0.9$ and the ambipolar diffusion coefficient of
$\eta_A=2.3\times 10^8$ m$^2$ s$^{-1}$. To initiate the instability we
perturbed the position of the interface by a multi-mode perturbation
containing 25 modes with wavelengths in the range $\lambda=10 \div 250$ km.
The resolution of the simulations was 1 km in each spatial direction. Such
high resolution is expected to be necessary to study non-ideal plasma
effects.

After subtracting the equilibrium conditions, we solve numerically the
quasi-MHD equations of conservation of mass, momentum, internal energy, and
the induction equation by means of our code {\sc mancha}
\citep{Felipe+etal2010, Khomenko+Collados2012} with the inclusion of the
physical ambipolar diffusion term in the equation of energy conservation and
in the induction equation. We evolve the ionization fraction in time by Saha
equation. Our code uses hyperdiffusivity for stabilizing the numerical
solution. To assure that the numerical diffusivity (whose action resembles
the physical diffusivity) does not affect the results of the simulations, we
kept the amplitude of artificial hyperdiffusive terms 2-3 orders of magnitude
lower that the physical ambipolar diffusion so that the characteristic time
scales of action of the former are orders of magnitude large. For each field
inclination we perform two identical simulations, one for the purely mhd case
and one with the ambipolar term switched on.

%%%%%%%%%%%%%%%%%%%%%%%%%%%%%%%%%%%%%%%%%%%%%%%%%%%%%%%%%%%%%%%%
\begin{figure*}
\center
\includegraphics[width=8.0cm]{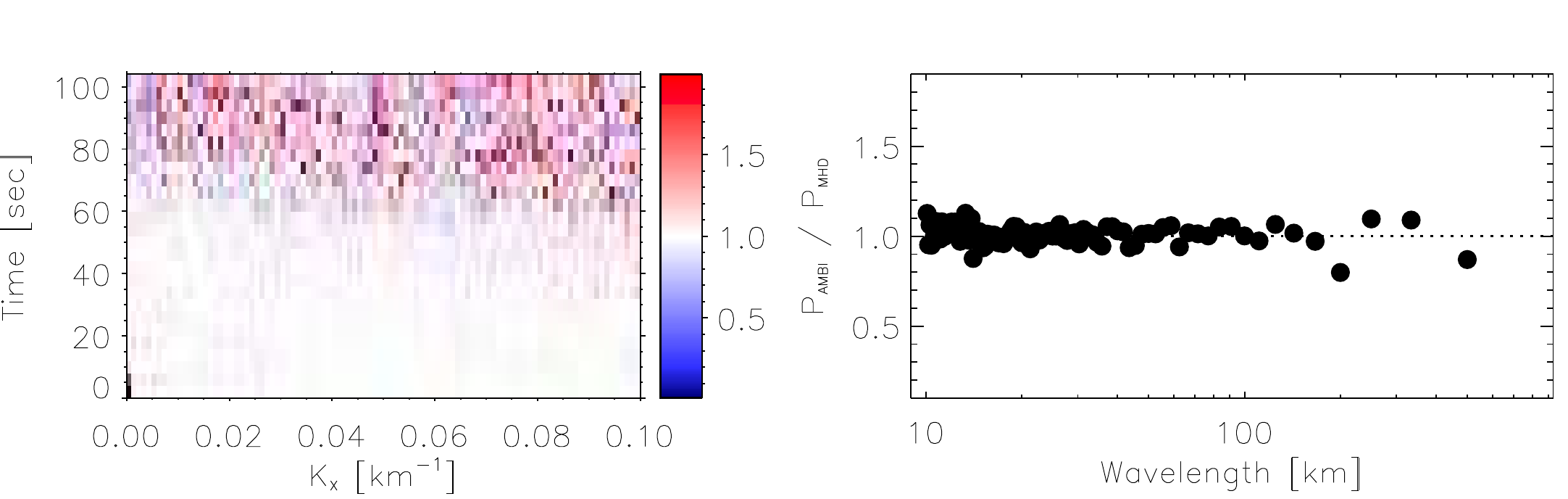}
\includegraphics[width=8.0cm]{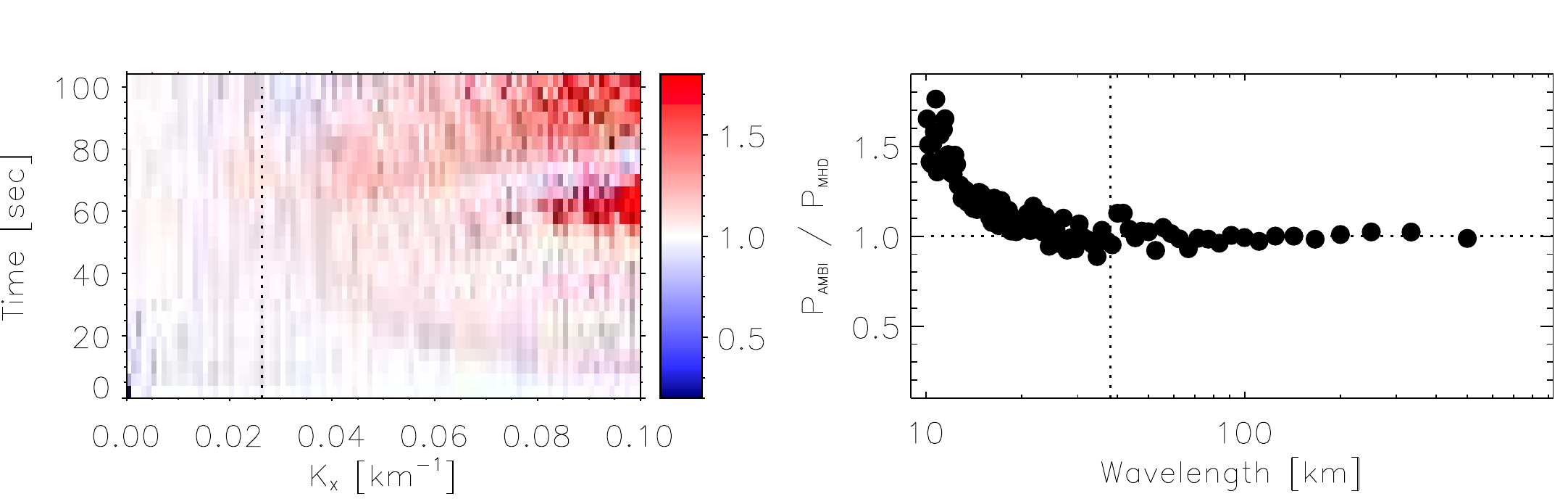}
\includegraphics[width=8.0cm]{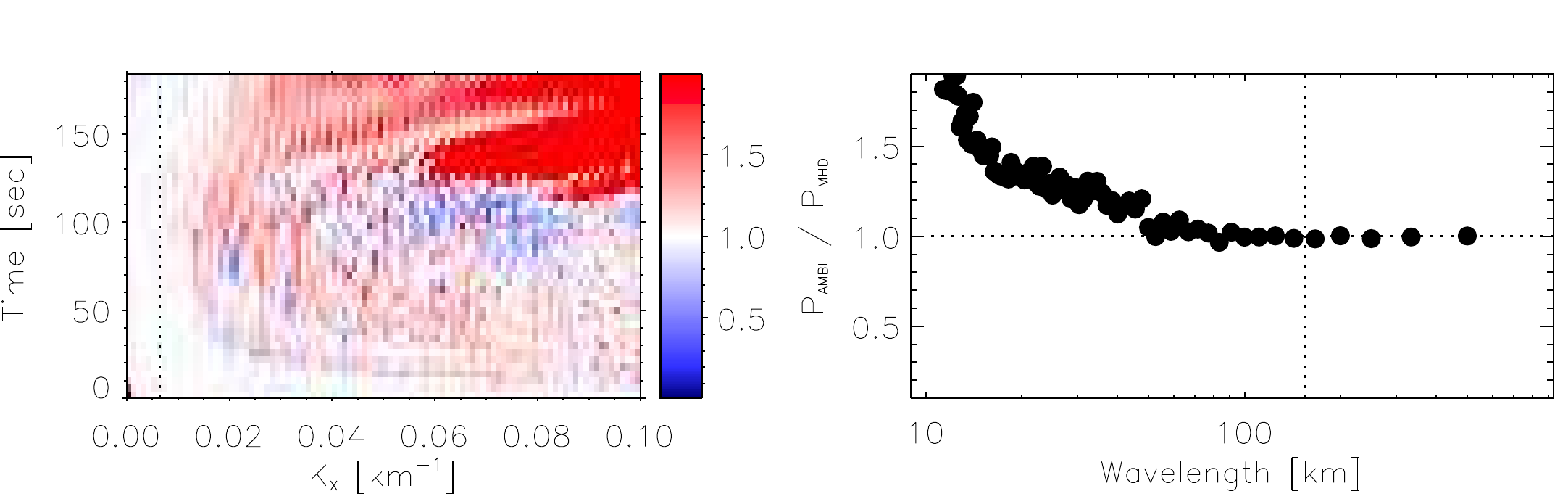}
\caption{Fourier analysis of scales developed in the simulations.  Left:
relative power of the ``ambipolar'' vs ``mhd'' simulation as a function of
horizontal wave number along the discontinuity, $k_x$, and time. Red colors
mean that the ``ambipolar'' simulation has more power. Right: time average of
the relative power from the panels on the left. Top: $\theta=90$\degree.
Middle: $\theta=89$\degree. Bottom: $\theta=88$\degree. In the last two
cases, the vertical dotted lines at the right panels marks the cut-off
wavelength $\lambda_c=38$ and $155$ km, correspondingly.} \label{fig:scales}
\end{figure*}
%%%%%%%%%%%%%%%%%%%%%%%%%%%%%%%%%%%%%%%%%%%%%%%%%%%%%%%%%%%%%%%%

Figure~\ref{fig:tevol} gives the time evolution of density perturbation in
the ``ambipolar'' simulations for the three field inclinations. The case of
the magnetic field normal to the plane of the instability
($\theta=90$\degree) is analogous to a purely hydrodynamical case since there
is not cut-off wavelength. Indeed, the simulation shows the development of
very small scales. The comparison of the ``mhd'' and ``ambipolar''
simulations (not shown in the Figure) reveals different (but statistically
equivalent, as is shown below) particular form of the turbulent flows, since
the ambipolar diffusion, acting on small scales, forces their different
evolution. The density variations have a pronounced asymmetry between the
large-scale upward rising bubbles and small-scale downflowing fingers, caused
by the mass conservation. Such asymmetric behavior can possibly explain the
preferred detection of the larger-scale upward rising bubbles in observations
\citep{Berger2010}.

By just rotating the field by 1\degree\ in $XY$ plane the scales developed in
the simulation are significantly changed (middle panels of
Fig.~\ref{fig:tevol}, $\theta=89$\degree), small scales disappear and only
few big drops are developed after about 200 sec of the simulation. The small
scales can not develop the instability because of the cut-off induced by the
magnetic field, $\lambda_c\approx 38$ km, for our equilibrium configuration
and $\theta=89$\degree. While not completely damped as in the purely mhd
case, the growth rate of small scales is still very low compared to large
scales, see the results by \cite{Diaz+etal2013}. One can appreciate from the
figure that at $t=40$ sec the dominant wavelength is around $\lambda\approx
 L/9$ km, while at $t\approx120$ sec it becomes $\lambda \approx L/6$ and finally after
$t\approx200$ sec the dominant wavelength is equal to $\lambda \approx L/3$,
being L=1 Mm the size of the simulation box. During the evolution of the
instability, the horizontal magnetic field component increases below and
above the drops as the field gets compressed by the flows. The increase of
the field produces a local increase of the cut-off wavelength, contributing
additionally to the damping of small scales. Moreover, small scales tend to
disappear with time due to non-linear interaction of harmonics with larger
scales, as was already demonstrated by other numerical works on the RTI,
\citep{Jun+etal1995b}.
Similar trend is observed in the simulation with $\theta=88$\degree, in the
bottom panels of Fig.~\ref{fig:tevol}. In this case, the critical wavelength
is $\lambda_c\approx 155$ km, and, accordingly, the developed scales are even
larger. Note that, at lower field inclinations, the evolution takes
progressively more time, since the tension force by the magnetic field slows
down the development of the instability, in agreement with the linear theory.

The velocity of flows lie in the range of 10--20 \kms, similar to
observations. The downflows are observed in about 2/3 of all points, and
upflows occupy the remaining 1/3 but the upflow velocities are, on average,
stronger. The comparison of the flows in the ``ambipolar'' and the ``mhd''
simulations in approximately linear stage of the RTI (first few tens of sec
of the simulation), reveals slightly larger velocities in the ``ambipolar''
simulations, i.e. this case is slightly more unstable.

To analyze the stability of different scales in the non-linear regime of the
RTI, at each time moment $t$ we calculated the power as a function of the
horizontal wave number $k_x$ by fourier-transforming in space a portion of
the snapshot of pressure variations around the discontinuity, and by
averaging the power for the vertical $k_z$ wave number to decrease the noise.
We then divided the ``ambipolar'' power map by the corresponding ``mhd'' map
for each pair of the simulations, $\theta=90$\degree, 89\degree\ and
88\degree. The result is given at the left panels of Fig.~\ref{fig:scales},
while the right panel shows the time-averages of the relative power maps.
In the case of the field directed normal to the perturbation plane
($\theta=90$\degree), we do not observe any significant change in power
between the ``ambipolar'' and ``mhd'' cases. The relative power fluctuates in
time, but the average remains around one for all harmonics (upper left
panel). This behavior is already anticipated from the linear analysis
\citep{Diaz+etal2013}. The relative power is significantly different to that
in the cases with $\theta=89$\degree\ and $88$\degree. Now there is a clear
increase in the growth rate of the small-scale harmonics in the ``ambipolar''
simulations compared to the ``mhd'' ones. The growth rate of the large-scale
harmonics is the same in both cases. The change in the behavior happens at
about $\lambda \sim 30$ km for the $\theta=89$\degree\ case, and at $\lambda
\sim 100$ km for the $\theta=88$\degree\ case, both numbers being close to
the corresponding cut-off wavelengths $\lambda_c$. This result confirms and
extends the conclusion from the linear theory that all RTI modes become
unstable when the presence of neutral atoms is accounted for in the analysis.
We obtain up to 50\% increase of the small-scale harmonics growth rate for
the $\theta=89$\degree\ case and up to 90\% increase for the
$\theta=88$\degree\ case. Our simulations show that partial ionization of
prominence plasma measurably influences its dynamics and must be taken into
account in future models.

\begin{acknowledgements}
This work contributes to the deliverables identified in FP7 European Research
Council grant agreement 277829, ``Magnetic connectivity through the Solar
Partially Ionized Atmosphere'', whose PI is  E. Khomenko.
\end{acknowledgements}

%\aareferences

\end{document}